\documentstyle[twoside,fleqn,espcrc2,graphicx]{article}
\newcommand{\eq}[1]{{\frenchspacing Eq.~(\ref{#1})}}
\newcommand{\fig}[1]{{\frenchspacing Fig.~(\ref{#1})}}
\newcommand{\tab}[1]{{\frenchspacing Table~(\ref{#1})}}

\newcommand{\beq}{\begin{equation}}
\newcommand{\eeq}{\end{equation}}

\newcommand{\AmS}{{\protect\the\textfont2\kern-.1667em\lower.5ex\hbox{M}\kern-.125emS}}
\title{A New Approach to $\eta'$ on the Lattice}

\author{K.\ Schilling \address{Dept.\ of Physics, University of Wuppertal,
42097 Wuppertal, Germany},
H.\ Neff \address{John von Neumann Institute for Computing
Research Center J\"ulich, 52425 J\"ulich, Germany},
N.\ Eicker $^{\rm a}$,
Th.\ Lippert $^{\rm a}$, 
J.\ W.\ Negele \address{Center for Theoretical Physics, MIT, Cambridge,
MA 02139, USA}}

\begin{document}
\begin{abstract}
  We perform an $\eta'$ mass analysis based on a total of 1130 dynamical gauge
  field configurations, with 5 different quark mass values on lattices of size
  $16^3 \times 32$ (SESAM) and $24^3 \times 40$ (T$\chi$L) at $\beta = 5.6$.
  We employ the stochastic estimator technique and spectral methods to deal
  with the disconnected piece of the flavour singlet correlation function. We
  demonstrate that very early plateau formation in the local $\eta'$ mass can
  be achieved by first ground state projecting the connected piece of its
  correlator.
\end{abstract}
\maketitle
\noindent           
{\it \hspace*{\fill} Am Fortschritt der  Moral beteiligt,\\}
{\it \hspace*{\fill} sind wir dar\"uber einig nun,\\}
{\it \hspace*{\fill} dass   nicht der Zweck die Mittel heiligt.\\}
{\it \hspace*{\fill} Doch der  Erfolg wird's ewig tun.\\[.2cm]}
{\small \hspace*{\fill} J.W. von Goethe }
\section{Introduction}
The masses of light non-singlet mesons have been determined quite accurately
in full QCD simulations.  The status of the $\eta'$ meson, however, is much
less satisfactory\cite{cppacs,ukqcd,sesam}, since it is affected for various
reasons by severe noise problems: (a) the $\eta'$-propagator requires the
evaluation of fermion loop operators, which so far has been achieved only with
stochastic estimator methods; (b) the Zweig-rule forbidden piece, which is a
loop-loop correlator, suffers from gauge field fluctuations; (c) from Wick
contraction the $\eta'$ propagator $C_{\eta'}(\Delta t)$ is computed as the
{\it numerical difference} between its connected and disconnected pieces,
leading to  loss of signal at large time separations $\Delta t$.

The challenge is to fight noise by exploiting the information from the
$\eta'$ correlator at {\it small time separations}.  In this
contribution, we intend to revisit the issue of the $\eta'$ signal by
testing the potential of a spectral representation.  The question is
whether in the current regime of full QCD calculations (we use SESAM
and T$\chi$L data), low eigenmodes are actually saturating the
fermion loop expressions.  Another concern is to improve the
groundstate projection in the $\eta'$-channel, aiming at very early
mass plateauing.

\begin{figure}[t]
\includegraphics[width=70mm]{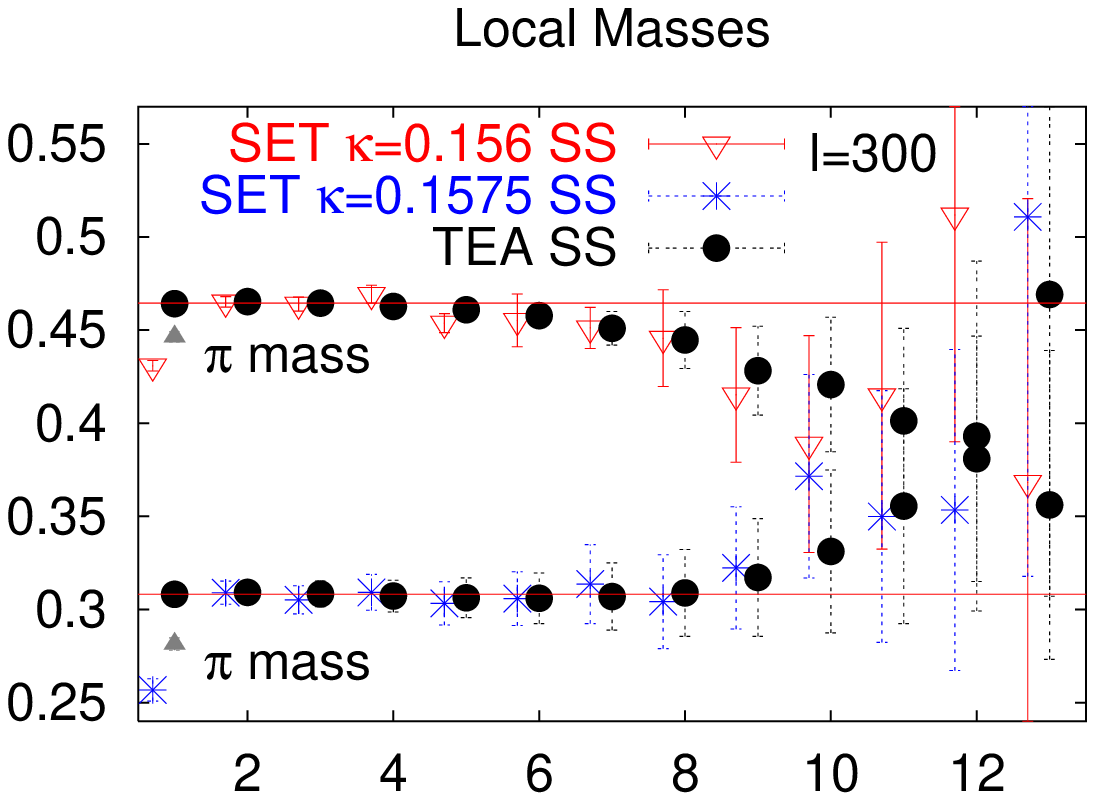}
\vskip-.3cm
\includegraphics[width=70mm]{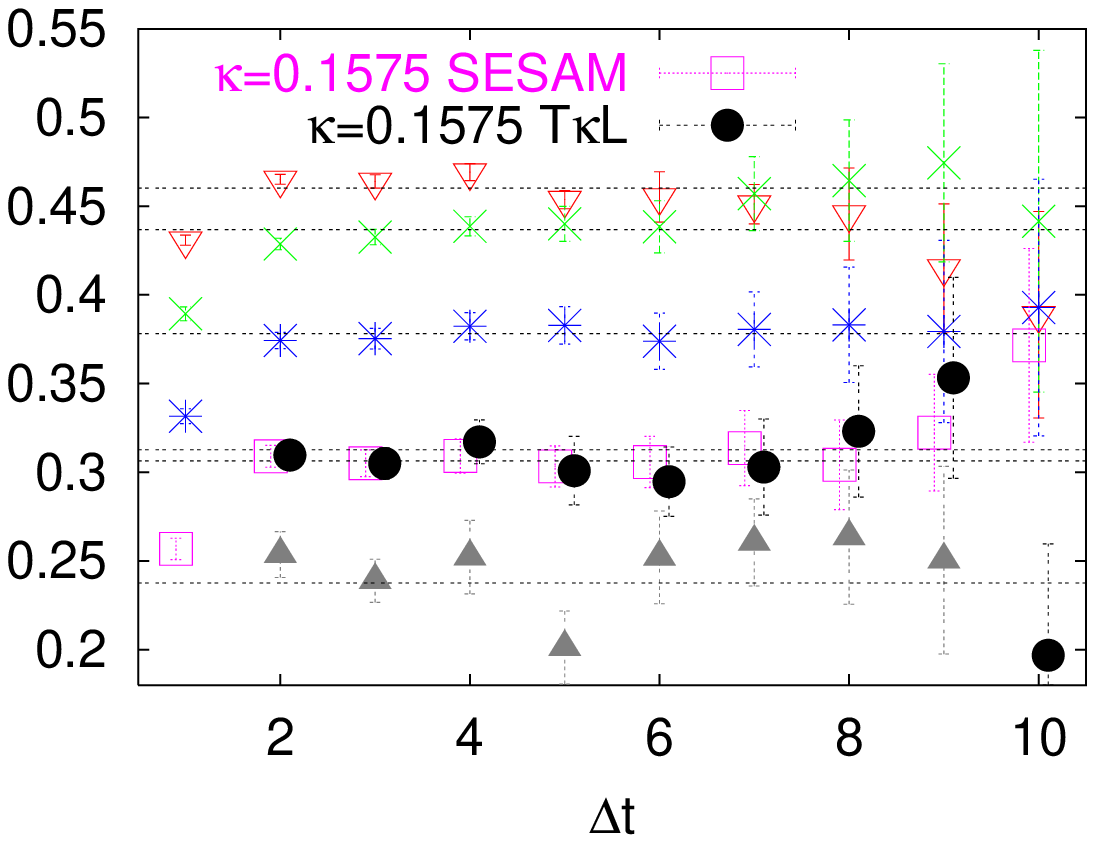}
\vspace{-10mm}
\caption{Upper frame: Comparison of mass plateaus from TEA and SET.  Lower
frame: Mass plateaus from SET, top down for $\kappa=0.156$ S, $0.1565$
S, $0.157$ S, $0.1575$ S and T, $0.158$ T, where S stands for SESAM and
T for T$\chi$L. The horizontal lines show the fitted mass values.}
\label{fig:fig5}
\vskip-0.2cm
\end{figure}

\section{The spectral approach}
Our truncated eigenmode approach (TEA)\footnote{For details see
\cite{tea}} applies the Lanczos method on the Hermitian form of
Wilson-Dirac matrix, $M$: $Q\equiv \gamma_5 M$ and computes the lowest
(in terms of their moduli) eigenmodes, $| \psi_i \rangle $: $ Q |
\psi_i \rangle = \lambda_i | \psi_i \rangle $.

In general we need 
the  spectral representation   of the quark propagator, $P$,
which implies the eigenfunctions
\beq
P(x,y)= \sum_{i = 1}^{V}\frac{1}{\lambda_i} 
\frac{ \gamma_5 | \psi_i (x) \rangle \langle \psi_i (y) | }
{ \langle \psi_i |  \psi_i \rangle}\; . \label{eq:inverse}
\eeq

We use the 300 lowest-lying eigenmodes, which at our lightest quark mass
amounts to about an equal computational effort as the use of 400 stochastic
sources.  For details of our analysis, see the parameters given in
\tab{tab:parameters}.

\begin{table}[t]
\caption{Parameter settings.  Smeared sources and sinks are
used\cite{smeared}. $N_{s/l}$ denotes the number of stochastic
sources/low eigenmodes and S (T) refers to SESAM (T$\chi$L)
lattices. lattices. The $eta'$ masses result from a single cosh-fit
over the $t$-range quoted with a jackknife analysis.}
\begin{tabular}{ccccc}
$\kappa_{sea}$ & $N_{s/l}$&\#confs& t-range  & $m_{\eta'}$ \\
\hline
SET:    &    &   &          &             \\ 
\hline
.1560/S & 400&195& 2-4      &$.4648(29)$ \\
.1565/S & 400&200& 2-6      &$.4326(39)$\\
.1570/S & 400&200& 2-10     &$.3775(72)$\\
.1575/S & 400&200& 2-8      &$.3071(87)$\\
.1575/T & 100&180& 2-8      &$.3068(69)$\\
.1580/T & 100&155& 2-8      &$.241(12)$\\
\hline
TEA:    &    &   &          &              \\
\hline
.1560/S & 300 &195& 1-4       &$.4645(28)$ \\
.1575/S & 300 &200& 1-8       &$.3080(70)$ \\
\end{tabular}
\label{tab:parameters}
\end{table}

The $\eta '$ propagator has the form
\begin{equation}
\label{eq:neff}
C_{\eta'}(\Delta t)={ C_{\pi}(\Delta t)}- 2 T(\Delta t) \; ,
\end{equation}
where the connected piece, $C_{\pi}$, is readily computed by use of linear
solvers.
The two-loop spectral expression, $T$,  reads:
\begin{eqnarray}\label{eq:tlf}
&&T(\Delta t) =  \nonumber \\
&&\sum_t \sum_{i,j} \frac{1}{\lambda_i}
\frac{ \langle \psi_i(t) | \psi_i(t) \rangle}
{ \langle \psi_i | \psi_i \rangle} 
\frac{1}{\lambda_j}
\frac{\langle \psi_j(t') | \psi_j(t') \rangle}
{\langle \psi_j | \psi_j \rangle}\; , 
\end{eqnarray}
with $t' = t + \Delta t$. 

 We propose to project out excited states from the connected piece by
the replacement $C_{\pi} \rightarrow C^g_{\pi}$, where $C^g_{\pi}$
stands for the fitted ground state correlation function \cite{tea}, and to
extract local $\eta'$ masses from the combination of `pseudodata'
\begin{equation}
\label{eq:neffs}
\tilde{C}_{\eta'}(\Delta t)={ C^g_{\pi}(\Delta t)}- 2 T(\Delta t) \; .
\end{equation}
Using \eq{eq:neffs} we obtain local mass plateaus that start at
very small values of $\Delta t$, as illustrated in \fig{fig:fig5},
where both TEA and stochastic estimator results \cite{sesam} are
shown.  Note that the statistical accuracy is sufficient to
discriminate the flavour non-singlet pseudoscalar mass very well from
the singlet one. We find the data from the TEA analysis to be less
noisy than the SET one. 

\begin{figure}[t]
\vskip-.9cm
\includegraphics[width=70mm]{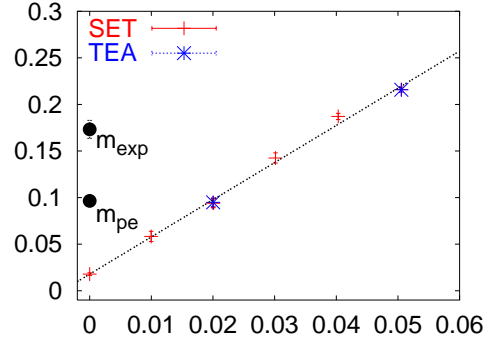}
\vspace{-10mm}
\caption{Linear chiral extrapolation of $m_{\eta'}^2$ in the quark
mass, $m_q$. The two flavor estimate in
\ref{eq:pseudomass} is denoted by $m_{pe}$. }
\label{fig:fig6}
\vskip-0.2cm
\end{figure}

Given the very early onset of the $\eta'$ mass plateaus, we can
perform single cosh fits to the flavour singlet correlator with safe
control on the underlying $t$-range.  The fit ranges in $t$ as well as
the $\eta'$-masses in lattice units for SET and TEA are listed in
\tab{tab:parameters}. We find perfect agreement between the results
from the SET and TEA analyses.
\section{Chiral extrapolation}
We translate the lattice mass values to physical units through the lattice
spacing $ a^{-1}_{\rho}(\kappa_{light}) = 2.302(64) \mbox{GeV} $ derived from
our previous light spectrum analysis~\cite{spectrum}. The critical and
physical light quark hopping parameters are $ \kappa_c=0.158507(44)$ and $
\kappa_{light} = 0.158462(42)$, respectively.

We perform the chiral extrapolation along $\kappa_{sea}=\kappa_{val}$,
assuming either $m_{\eta'}$ or $m^2_{\eta'}$ to be linear in the quark
mass.  Our fits favour the dependence $m^2 = c + c'm_q$ 
from the $\chi^2/d.o.f.$ in \tab{tab:eta_masses}.

To compare with the physical $\eta'$ mass, we have to  transform  the
mass from the real world of three flavours into our $N_f = 2$ setting.
 This can be done by making use of the experimental mass splitting
between flavour singlet and non-singlet states
\beq
M^2_{0;N_f=3} = M^2_{\eta';N_f=3} - \left( 2 M^2_K -M^2_{\eta} \right)  
\eeq 
and the Witten-Veneziano formula\footnote{for a recent discussion
on the lattice version of this eq., see \cite{rossi}} 
\beq 
M^2_0 = 2
N_f \chi / F^2_{\pi}\; .  
\eeq 
For $N_f=2$ all non-singlet masses are
degenerate. Hence the above equations translate to 
\beq
M^2_{\eta';N_f=2} = 2/3 M^2_{0;N_f=3} + M^2_{\pi} \;. \label{eq:wveta} 
\eeq 
Inserting the physical mass values on the
right hand side of \eq{eq:wveta} leads to the two flavour $\eta'$ mass
\beq M_{\eta';N_f=2} = 715 \mbox{MeV} \; .
\label{eq:pseudomass}
\eeq

By inspection of \tab{tab:eta_masses} we find that (at our lattice spacing!)
the linear (in quark mass) chiral extrapolations of both $m_{\eta'}$ and
$m^2_{\eta'}$ are definitely above the pion mass, yet significantly
below the two-flavour pseudoexperimental value estimated in \eq{eq:pseudomass}.

\begin{table}
\caption{Chiral extrapolation of the lattice $\eta'$ masses. The
errors stem from a $\chi$ square fit to the measured $\eta'$ masses.}
\label{tab:eta_masses}
\begin{tabular}{cccc}
Fit & $m_{\eta'}$ & $M_{\eta'}$[MeV] & $\chi^2/d.o.f.$ \\
\hline
$m$-fit  & 0.214(7) & 493(30)  & 5.6 \\
$m^2$-fit & 0.138(15) &  318(43) & 2.2 \\
\hline
\end{tabular}
\end{table}

\section{Summary and Conclusion}
We have presented a new approach to the computation of flavour singlet masses,
where the two-loop correlators are estimated by a spectral representation. We
have shown that ${\cal O}(300)$ low eigenmodes suffice to saturate the fermion
loop contribution to the $\eta'$ propagator in the regime of quark masses of
the SESAM and T$\chi$L settings.

We found that, in both the TEA and SET analyses, most of the excited
state contamination to the $\eta'$ propagator is related to its
connected contribution: by a conventional ground state projection of
the latter, one can achieve a strikingly early onset of plateau
behaviour in the effective $\eta'$ mass. This then leads to 14 \%
statistical accuracy of $M_{\eta'}$ after chiral extrapolation.

In our setting, TEA and SET are equally cost effective.  It is obvious
that TEA is bound to win as we shall go to lighter quark masses in the
future; there iterative solvers -- unlike the Lanczos procedure --
will suffer severe convergence problems due to large condition
numbers, while eigenmode expansions will converge even better.

It appears worthwhile to investigate the potential of TEA for
disconnected matrix elements. Research in this direction is under way.

\end{document}